\documentclass[aps,pre,twocolumn,floatfix,10pt,showpacs,superscriptaddress]{revtex4-1}

\usepackage{graphicx}
\usepackage{verbatim}
\usepackage{comment}
\usepackage{amssymb}
\usepackage{amsmath}
\usepackage{ mathrsfs }
\usepackage{enumerate}
\usepackage{psfrag}
\usepackage{pstricks}
\usepackage{color}
\usepackage{hyperref}
\usepackage{enumitem}
\emergencystretch=\maxdimen
\definecolor{ao}{rgb}{0.0, 0.5, 0.0}
\newcommand{\lsim}{\mathrel{\hbox{\rlap{\lower.55ex \hbox{$\sim$}} \kern-.3em \raise.4ex \hbox{$<$}}}}

\begin{document}

\title{Lanczos Boosted Numerical Linked-Cluster Expansion for Quantum Lattice Models}

\author{Krishnakumar Bhattaram}
\affiliation{Lynbrook High School, 1280 Johnson Ave, San Jose, CA 95129, USA}
\affiliation{Department of Physics and Astronomy, San Jose State
University, San Jose, CA 95192, USA}
\author{Ehsan Khatami}
\affiliation{Department of Physics and Astronomy, San Jose State
University, San Jose, CA 95192, USA}

\begin{abstract}
Numerical linked-cluster expansions allow one to calculate finite-temperature properties of quantum 
lattice models directly in the thermodynamic limit through exact solutions of small clusters. However, 
full diagonalization is often the limiting factor for these calculations. Here, we show that a partial 
diagonalization of the largest clusters in the expansion using the Lanczos algorithm can be as useful
as full diagonalization for the method while mitigating some of the time and memory issues. 
As a test case, we consider a frustrated Heisenberg model on the checkerboard lattice
and find that our approach can lead to much more efficient calculations in a parallel environment.
\end{abstract}

\maketitle

%%%%%%%%%%%%%%%%%%%%%%%%%%%%%%%%%%%%%%%%%%%%%%%%%%%%%%%%
% INTRODUCTION
%%%%%%%%%%%%%%%%%%%%%%%%%%%%%%%%%%%%%%%%%%%%%%%%%%%%%%%%

\section{Introduction}

Over the past decade, the numerical linked-cluster expansion (NLCE)~\cite{M_rigol_06,M_rigol_07a,M_rigol_07b} has become
a versatile and powerful technique for solving quantum lattice models and studying finite-temperature
properties of strongly-correlated systems in the thermodynamic limit. So far, it 
has seen applications to a variety of problems including magnetic models~\cite{M_rigol_07a,M_rigol_07c,E_khatami_11,e_khatami_11c},
itinerant electron models on several different geometries~\cite{M_rigol_07b,E_khatami_11b,b_tang_12,e_khatami_16}, 
quantum quenches in the thermodynamic limit~\cite{m_rigol_14,b_tang_15}, 
entanglement~\cite{a_kallin_13,a_kallin_14,e_stoudenmire_14}, and even the hard problem of dynamics in the thermodynamic limit
for system in or out of equilibrium~\cite{k_mallayya_18,i_white_17,m_nichols_18}.

At the heart of the method lies the exact solution of model Hamiltonians on relatively small clusters, which have
open boundaries and  unusual topologies, using exact diagonalization. This step is the most 
computationally expensive and therefore the limiting factor of the method. For models such as the Fermi-Hubbard model,
the growth in the size of the Hilbert space with cluster size puts the calculations quickly up against an 
exponential wall. In other cases, the factorially growing number of clusters that need to be diagonalized is the limiting factor.

In cases where only the ground state is of interest, and in the absence of long-range entanglement, which is unfavorable to the 
convergence of the series, other techniques such as the Lanczos algorithm or the density matrix renormalization 
group have been employed~\cite{e_khatami_11c,e_stoudenmire_14}.

The main idea behind our study is to employ the Lanczos technique to obtain not only the ground state, but 
also a part of the low-energy spectrum of the model in order to access nonzero, but low, temperatures for 
large clusters in the NLCE.
The goal is to combine the partial-diagonalization for solving clusters in the last order of the series
with the traditional full diagonalization of smaller cluster at lower orders to push the convergence beyond 
temperatures that have been accessible before. The larger the clusters at higher orders, the lower the
temperatures they contribute to in the NLCE, and the smaller the percentage of the lowest-lying 
eigenenergies  that is needed.

We benchmark our results from this technique against those obtained from full diagonalization in the 
NLCE for the antiferromagnetic Heisenberg model on the checkerboard lattice
with frustration. We find that in a massively parallel environment, the Lanczos-based method even with 
the very expensive full re-orthogonalization step can offer a large speedup  over the traditional diagonalization method.

The exposition is as follows. In Sec.~\ref{sec:model}, we introduce the model. 
In Sec.~\ref{sec:method}, we provide the basics of the NLCE and discuss how it can benefit from the 
Lanczos algorithm. We discuss the results for the model, scaling and some details of the parallelization scheme 
in Sec.~\ref{sec:results}

%%%%%%%%%%%%%%%%%%%%%%%%%%%%%%%%%%%%%%%%%%%%%%%%%%%%%%%%
% Model
%%%%%%%%%%%%%%%%%%%%%%%%%%%%%%%%%%%%%%%%%%%%%%%%%%%%%%%%

\section{Checkerboard Lattice Heisenberg Model}
\label{sec:model}

\begin{figure}[h!]
\includegraphics[width=1in]{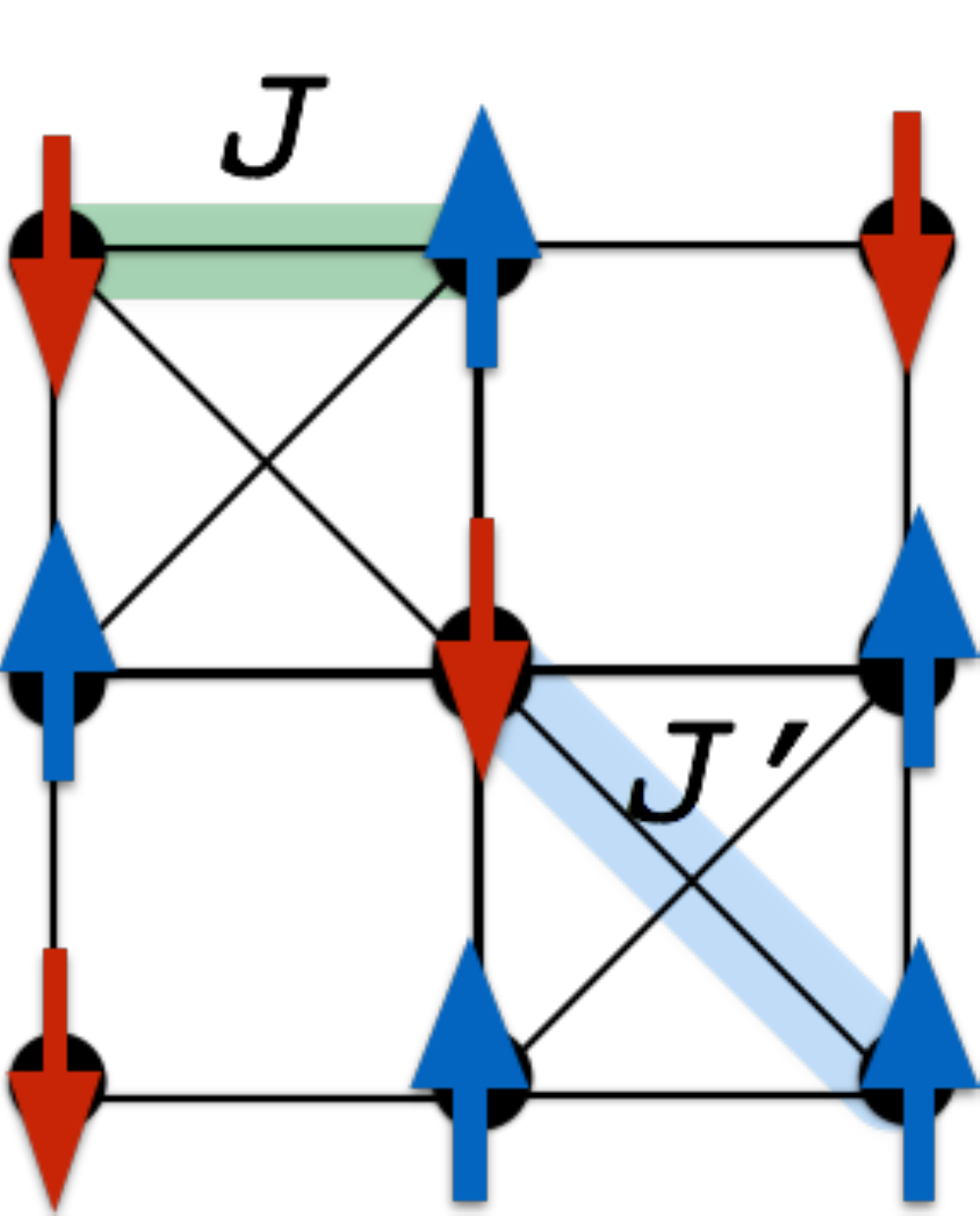}
\caption{A $3 \times 3$ section of the checkerboard lattice Heisenberg model. 
$J$ is the strength of the exchange interaction on nearest neighbor bonds and $J'$ is the interaction on 
next nearest neighbor bonds in every other $2\times 2$ plaquette.
\label{fig:lattice}}
\end{figure}

The Hamiltonian we consider here is the spin-1/2 quantum Heisenberg model on the checkerboard lattice, which is written as 
\begin{equation}
\label{eq:Heis}
\hat{H} = \sum\limits_{ij} J_{ij} {\bf \hat{S}}_i\cdot {\bf \hat{S}}_j,
\end{equation}
where ${\bf \hat{S}}_i$ is the spin-1/2 vector at site $i$, and $J_{ij}=J$ for nearest neighbor
$i$ and $j$ and $J_{ij}=J'$ for next nearest neighbor $i$ and $j$ on every other plaquette 
in a checkerboard pattern as shown in Fig.~\ref{fig:lattice}. $J'=0$ represents the square 
lattice Heisenberg model. Here, we consider the test case $J=0.5$ and $J'=1.0$. So,
$J'$  sets the unit of the energy throughout the paper

%%%%%%%%%%%%%%%%%%%%%%%%%%%%%%%%%%%%%%%%%%%%%%%%%%%%%%%%
% NLCE
%%%%%%%%%%%%%%%%%%%%%%%%%%%%%%%%%%%%%%%%%%%%%%%%%%%%%%%%

\section{Numerical Linked-Cluster Expansion}
\label{sec:method}

In the numerical linked-cluster expansion a given extensive property $P(\mathscr{L})$ of the lattice model
is expressed as a sum over the contributions to that property from every cluster that can be embedded in 
the lattice $\mathscr{L}$. While a version of the method can be employed for finite system sizes, its main
advantage  has been in its applications to the models in the thermodynamic limit,
$\mathscr{L} \to \infty$. In that limit, the series expansion is given as:
\begin{equation}
\label{eq:nlce_exp}
P(\mathscr{L})/\mathscr{L}= \sum_{c} L(c) W_P(c)
\end{equation}
where $c$ are topologically-distinct clusters that can be embedded in $\mathscr{L}$, 
$L(c)$ is the number of ways per site cluster $c$ can be embedded in the lattice, 
and $W_P(c)$ is the contribution to property $P$ computed via the inclusion-exclusion principle:
\begin{equation}
\label{eq:nlce_weight}
W_P(c) = P(c) - \sum\limits_{s \subset c}  W_P(s),
\end{equation}
where $s$ is a cluster that can be embedded in $c$ (a sub-cluster of $c$), and $P(c)$ is the property 
calculated for cluster $c$ using exact diagonalization (ED). $L(c)$ can be thought of as the number of point 
group symmetry operations for the underlying lattice that give the cluster a distinct orientation.
More information on how $L(c)$ are calculated and other details of the algorithm can be found 
in Ref.~\cite{b_tang_13b}.

The calculation of properties $P(c)$ using ED is the most computationally expensive part
of the NLCEs. For example, in a site expansion, for which order $l$ of the series includes
all clusters up to $l$ sites, calculations for the square or honeycomb lattice Fermi-Hubbard models
have so far been limited to $l=9$~\cite{E_khatami_11b,e_khatami_12b,b_tang_12,b_tang_13}. 
Even though there are only 112 topologically-distinct clusters to diagonalize for the square 
lattice in the 9th order in that case, the calculations are limited by memory and time requirements for 
diagonalization of the largest matrices for clusters with a size larger than 9 sites. For quantum magnetic models in which
the size of the Hilbert space ($n_{Hilbert}$) grows significantly slower with the system size than for Fermi-Hubbard 
models, the calculations have been limited to about 15 orders on the square lattice~\cite{b_tang_13b,b_tang_15b}, 
not by memory requirements, but by the large number of clusters that need to be solved in higher orders. 

\begin{figure}[t]
\includegraphics[width=3.3in]{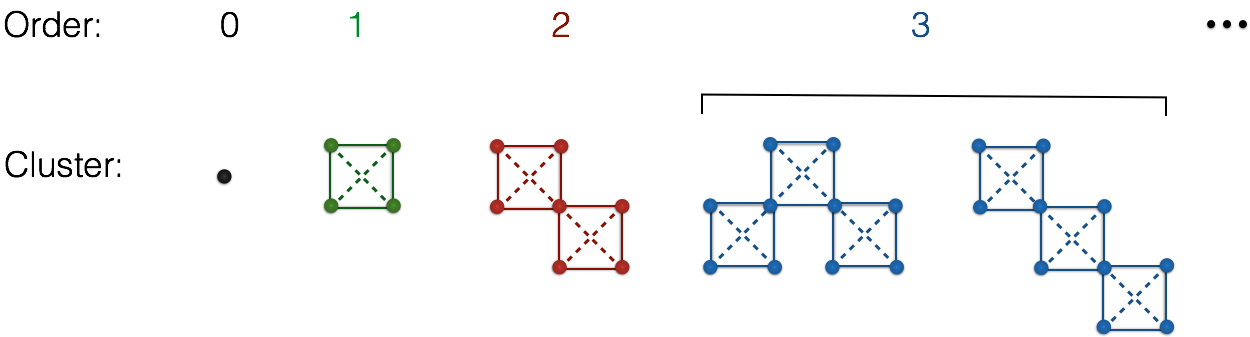}
\caption{Square expansion for the checkerboard lattice Heisenberg model, where clusters at higher orders are generated by 
adding corner-sharing $2\times 2$ blocks with next nearest neighbor bonds to smaller clusters in lower orders.
Shown are the topologically distinct clusters up to the third order of the expansion.
\label{fig:clusters}}
\end{figure}

In those cases, one can put the limitation back on the ED and push the convergence to lower
temperatures by employing expansion schemes that access much larger clusters more quickly as the
order increases. That is accomplished through increasing the size of the building block 
used to generate clusters. For example, an expansion 
with squares as building blocks was used to study the checkerboard and square lattice Heisenberg models
with clusters up to 19 sites in only 6 orders~\cite{E_khatami_11}. Figure~\ref{fig:clusters} shows the 
first five topologically-distinct clusters in the square expansion. In previous studies, triangular building blocks were 
used to study magnetic models on the Kagome lattice~\cite{e_khatami_11c,e_khatami_12a}, and 
rectangular clusters were considered in the study of entanglement at a two-dimensional critical point~\cite{a_kallin_13}.

Here, we aim to address, to the extent possible, the limitation on time and memory for 
full diagonalization, which remains the main issue in the NLCE. 
We combine the square expansion NLCE with an efficient partial diagonalization 
offered by the Lanczos algorithm to access low temperatures much faster than previously possible. We 
base our approach on a key observation that only small clusters in low orders of the expansion
contribute to properties at the highest temperatures; the inclusion of larger clusters in the series
at higher orders does not change the results at high energies/temperatures. Therefore, a 
partial knowledge of the eigenvalues and corresponding eigenvectors at the low-energy 
end of the spectrum may be sufficient to deduce new information from larger clusters

%%%%%%%%%%%%%%%%%%%%%%%%%%%%%%%%%%%%%%%%%%%%%%%%%%%%%%%%
% Lanczos
%%%%%%%%%%%%%%%%%%%%%%%%%%%%%%%%%%%%%%%%%%%%%%%%%%%%%%%%

\subsection{Efficiency of the Lanczos Algorithm for the NLCE}
The Lanczos algorithm~\cite{c_lanczos_50} has had profound applications in solving discrete quantum systems, 
notably when the ground state is of interest. 
The following section will give an overview of the Lanczos method in terms of its applicability to the NLCE. For further
information concerning the mechanics and derivation of the Lanczos algorithm, see Ref.~\cite{c_lanczos_50}

The algorithm in its most basic form is an iterative Krylov subspace method in which an orthogonal basis is built and 
approximations of eigenvectors are found by projecting our matrix onto the subspace. 
It is well known for its superior space and time efficiency in solving for the ground state of very large systems. 
However, the method is known to have numerical instabilities in finite-precision mathematics~\cite{c_paige_72}, which manifest 
themselves in a loss of orthogonality 
of the Lanczos vectors which define the subspace. This leads to spurious degeneracies of the largest-magnitude eigenvalues, 
which means that while the ground-state behavior may be fully captured, finite-temperature properties are very difficult to ascertain.

\begin{figure}[t]
\includegraphics[width=3.3in]{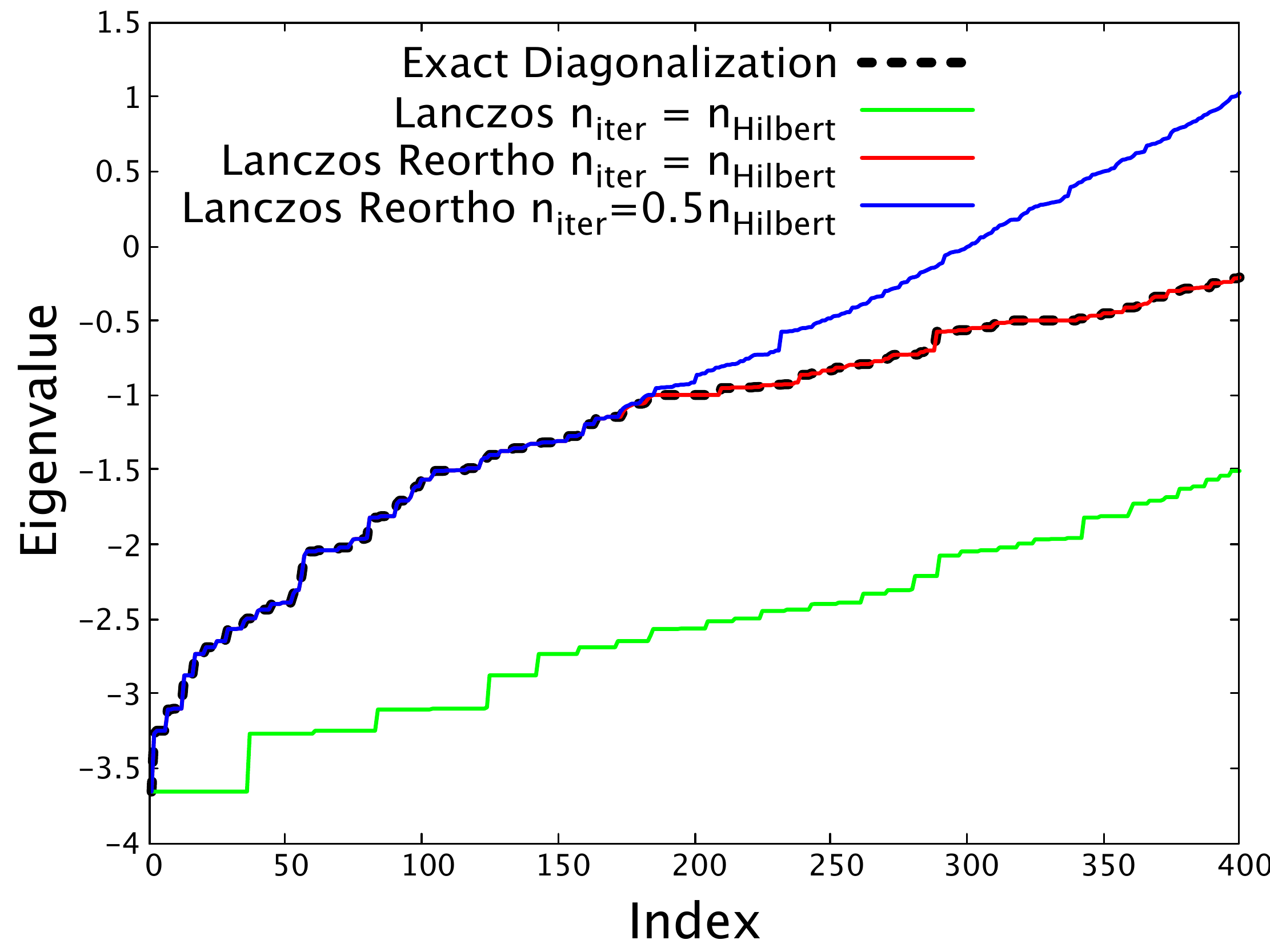}
\caption{Sorted first 400 eigenvalues of the model Hamiltonian for a 16-site cluster in the four spin-up sector with $n_{Hilbert} = 1820$.
Results from the Lanczos algorithm deviate from the exact ones very early on in the spectrum due to the existence of
degeneracies and the loss of orthogonalization between vectors in the Krylov space. A full re-orthogonalization step
in the algorithm restores orthogonality of the Lanczos vector at every iteration and 
captures a larger percentage of the exact spectrum as the number of iteration increases.
\label{fig:eigval}}
\end{figure}

The solutions to maintaining orthogonality are varied~\cite{h_simon_84},
however, the most basic in capturing the whole set of eigenvalues to high precision is the application at every iteration of a 
Gram-Schmidt orthogonalization with all previous vectors, a full reorthogonalization scheme. Figure~\ref{fig:eigval} shows the 
exact eigenvalues of the Hamiltonian for a 16-site cluster in the spin sector with four spin-ups and twelve spin-downs
compared with the values from the Lanczos method with and without full reorthogonalization and two different numbers of iterations ($n_{iter}$). 
It is expected that with $n_{iter}=n_{Hilbert}$ the Lanczos algorithm can find most (but not all) eigenvalues with high degree of precision. 
The severe degeneracy of the lowest eigenvalues in the basic Lanczos method and its resolution after reorthogonalization can be seen in the figure. 

The reorthogonalization process requires the storage of about $n_{iter}$ Lanczos vectors, which has similar 
memory requirements to storing the Hamiltonian matrix when $n_{iter}\sim n_{Hilbert}$, and also
 greatly increases the time complexity of the process due to a $n_{iter}^2\times n_{Hilbert}$ scaling. The question thus raised is whether or not 
the reorthogonalized Lanczos scheme is in fact advantageous over exact diagonalization. Figure~\ref{fig:eigval} includes a plot 
for the case where the number of iterations is half the size of the Hilbert space. It can be seen that a large fraction of low-energy states
have been captured with high accuracy in that case. A key feature of the NCLE is that with the first few orders, the series
is already convergent at high temperatures where correlations in the system are short ranged. Larger clusters 
in higher orders provide information for properties 
in the thermodynamic limit only at lower temperatures where correlations grow larger. 
On the other hand, the Lanczos algorithm can provide very accurate information about the 
low-temperature behavior of the largest clusters in the series even when the number of
iterations is less than the size of the Hilbert space as low-lying eigenvalues can be 
converged, leading to much smaller memory and time requirements in comparison with ED. 
In other words, as system sizes increase by increasing the order and the region of convergence of the series is pushed to 
lower temperatures, one can also expect that fewer iterations are necessary. 
Furthermore, the Lanczos algorithm is readily parallelizable both in the Hamiltonian multiplication and in the reorthogonalization
step in order to further reduce the diagonalization time. A discussion about scaling is provided in Sec.~\ref{sec:scaling}.

%%%%%%%%%%%%%%%%%%%%%%%%%%%%%%%%%%%%%%%%%%%%%%%%%%%%%%%%
% RESULTS
%%%%%%%%%%%%%%%%%%%%%%%%%%%%%%%%%%%%%%%%%%%%%%%%%%%%%%%%

\section{Results}
\label{sec:results}

\begin{figure}[t]
\includegraphics[width=3.4in]{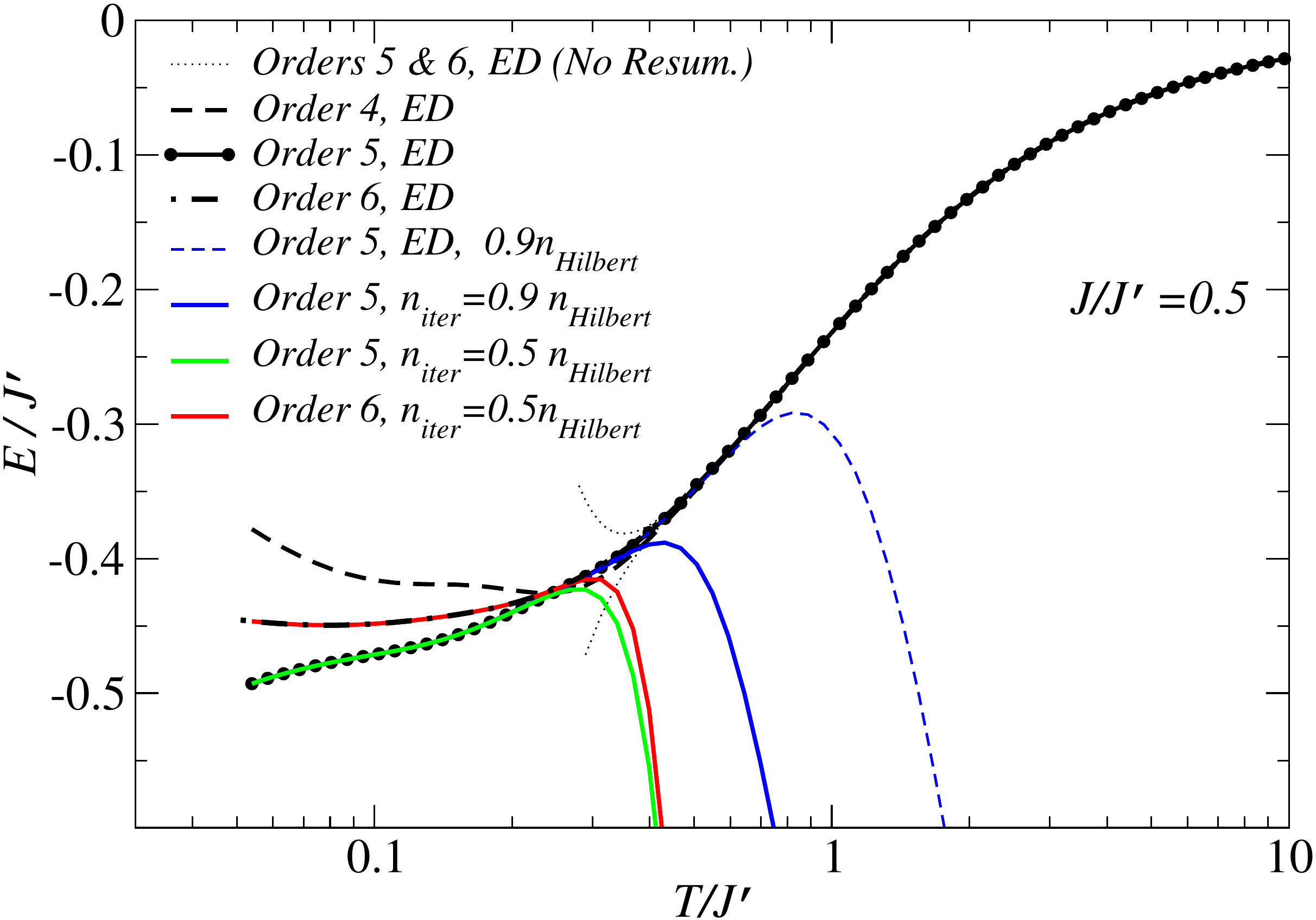}
\caption{Energy per site of the Heisenberg model on a checkerboard lattice with $J=0.5$ and $J'=1.0$ as a 
function of temperature. The results within the convergence region are valid in the thermodynamic limit. Thin 
dotted lines are bare sums in the $5$th and $6$th orders. The latter is taken from Ref.~\cite{E_khatami_11}. Thick
black lines are results in the $4$th, $5$th, and $6$th orders from Euler resummations after regular sum of the first three 
terms in the series. These results are obtained through full diagonalization using a Lapack routine.
Color lines show the results from our Lanczos algorithm for the $5$th or $6$th orders with different number of iterations. 
As the latter increases, a larger percentage of the low-lying eigenvalues for every cluster in that 
order approach their exact values and the corresponding curves match the exact curve 
up to higher temperatures. Blue dashed line represents results from ED for the case where 90\%
of the eigenvalues are used in the thermal averages.
\label{fig:energy}}
\end{figure}

To benchmark the performance of the Lanczos-boosted NCLE, we start with the energy per site of our 
Heisenberg model with $J=0.5$ and $J'=1.0$ calculated up to the sixth order of the square expansion 
NLCE (see Fig.~\ref{fig:energy}). We use Intel's math kernel library (MKL) for our ED as well as our Lanczos 
method with full reorthogonalization. ED results for the sixth order are taken from Ref.~\cite{E_khatami_11} where 
ScaLapack, a distributed-memory version of the linear algebra package (Lapack), was used for diagonalization 
to meet the enormous memory and time demands. We have used the Lanczos diagonalization for clusters only in the last order 
of the expansion while other smaller clusters were solved using ED. We have also used ED to diagonalize matrices
smaller than 2000 in linear size in the last order to avoid issues in Lanczos that may arise with small matrices. 
We have used the Euler method for numerical resummation of the series~\cite{M_rigol_07a,b_tang_13b}. 
Results for the fifth and sixth orders before the resummations are also shown as thin dotted lines. Results 
from another resummation method agree with the outcome of the Euler resummation in its region of convergence.

We find that Lanczos with $n_{iter}=0.5n_{Hilbert}$ is sufficient to reach the lowest convergence 
temperature of the series with five or six orders. The latter are shown as green and red curves in Fig.~\ref{fig:energy}.
One can see that they perfectly agree with results from ED up to temperatures around $0.3$, slightly above
the temperature at which the convergence of the series with six orders is lost. 
The exact energies in the thermodynamic limit beyond this point are already accessible to lower orders. 
We also show that by increasing $n_{iter}$ one can systematically reach higher temperatures. With $n_{iter}=0.9n_{Hilert}$
the agreement with exact results extends to $T\sim 0.4$. 

As noted above, $n_{iter}$ does not represent the number of exact eigenvalues obtained in the Lanczos 
algorithm. We demonstrate that by using the smallest 90\% of exact eigenvalues from ED in the Boltzmann average 
of the energy in the fifth order, which results in the blue dashed curve in Fig.~\ref{fig:energy},
surpassing the performance of the Lanczos algorithm with  $n_{iter}=0.9n_{Hilert}$.

\begin{figure}[t]
\includegraphics[width=3.3in]{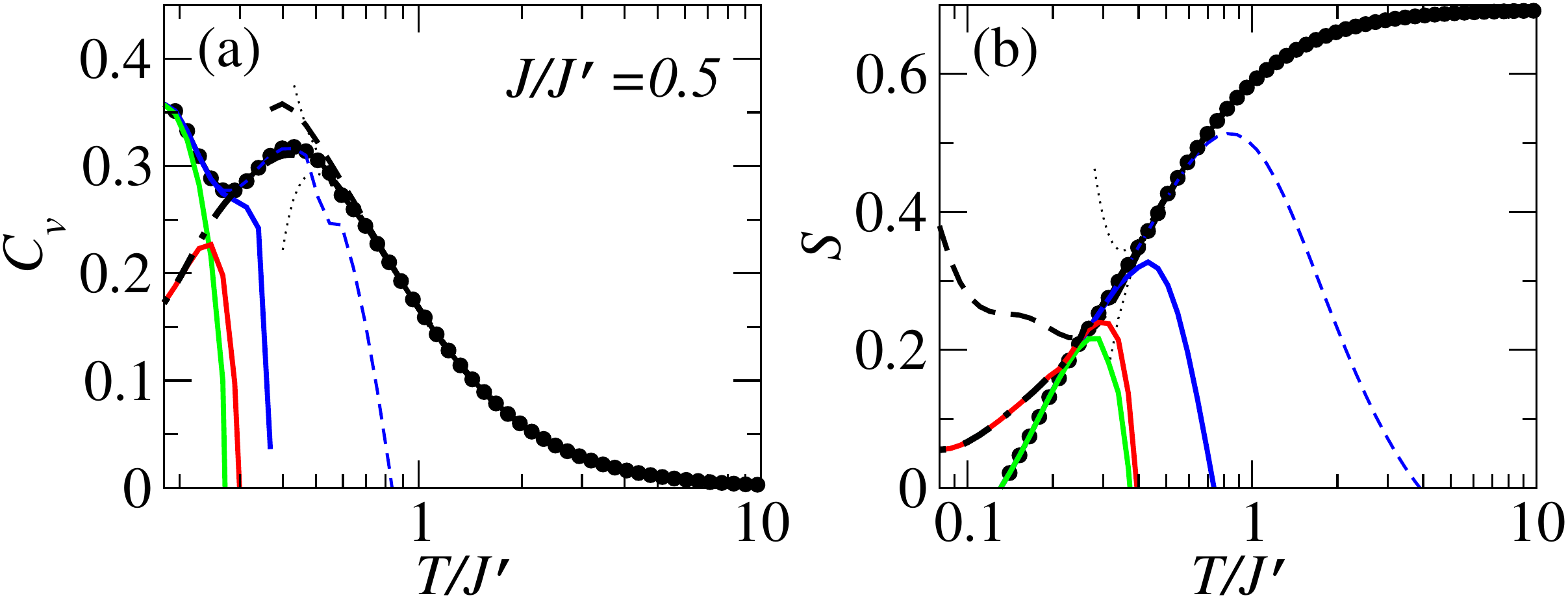}
\caption{(a) The heat capacity, and (b) entropy of the Heisenberg model on a checkerboard lattice with $J=0.5$ and $J'=1.0$
as a function of temperature. Lines are the same as Fig.~\ref{fig:energy}. In case of the heat capacity, partial diagonalization using 
the Lanczos algorithm barely reaches the minimum convergence temperature of the series with 6 orders. However, a $n_{iter}$ of
50\% of the size of the Hilbert space is enough to reach the convergence temperature for the entropy with $5$ or $6$ orders.
\label{fig:pro}}
\end{figure}

As the order increases and the convergent region is extended to lower temperatures, 
the number of iterations can be increasingly smaller fractions of $n_{Hilbert}$, leading to the superiority in both space
and time complexity of the Lanczos-assisted diagonalization over ED. This is due to the ability of the  
Lanczos method to capture the lowest eigenvalues with the least number of iterations.
The calculation of properties in the sixth order with clusters up to 19 sites was only possible previously
through the use of Scalapack routines as done in Ref.~\cite{E_khatami_11}. 
Here, we have obtained results for the sixth order with $n_{iter}=0.5n_{Hilert}$ in significantly less time.
As one can see in Fig.~\ref{fig:energy}, those results already capture the average energy before the 
series loses convergence around $T=0.3$.

The efficiency of our approach in using the Lanczos algorithm for treating clusters in the last oder of the NLCE 
depends on the property of interest, and most likely, on the model under investigation.
In Fig.~\ref{fig:pro}(a), we show that using half as many iteration in Lanczos as
the size of the Hilbert space is insufficient to capture the heat capacity of the same system studied in Fig.~\ref{fig:energy}
around the lowest convergence temperature of the series with up to six orders. The same is not true for the
entropy, shown in Fig.~\ref{fig:pro}(b), where we observe a behavior similar to that of the energy in terms of 
temperature access with the Lanczos algorithm.

%%%%%%%%%%%%%%%%%%%%%%%%%%%%%%%%%%%%%%%%%%%%%%%%%%%%%%%%
% SCALING
%%%%%%%%%%%%%%%%%%%%%%%%%%%%%%%%%%%%%%%%%%%%%%%%%%%%%%%%

\subsection{Parallelization and Scaling}
\label{sec:scaling}

\begin{figure}[t]
\includegraphics[width=3.3in]{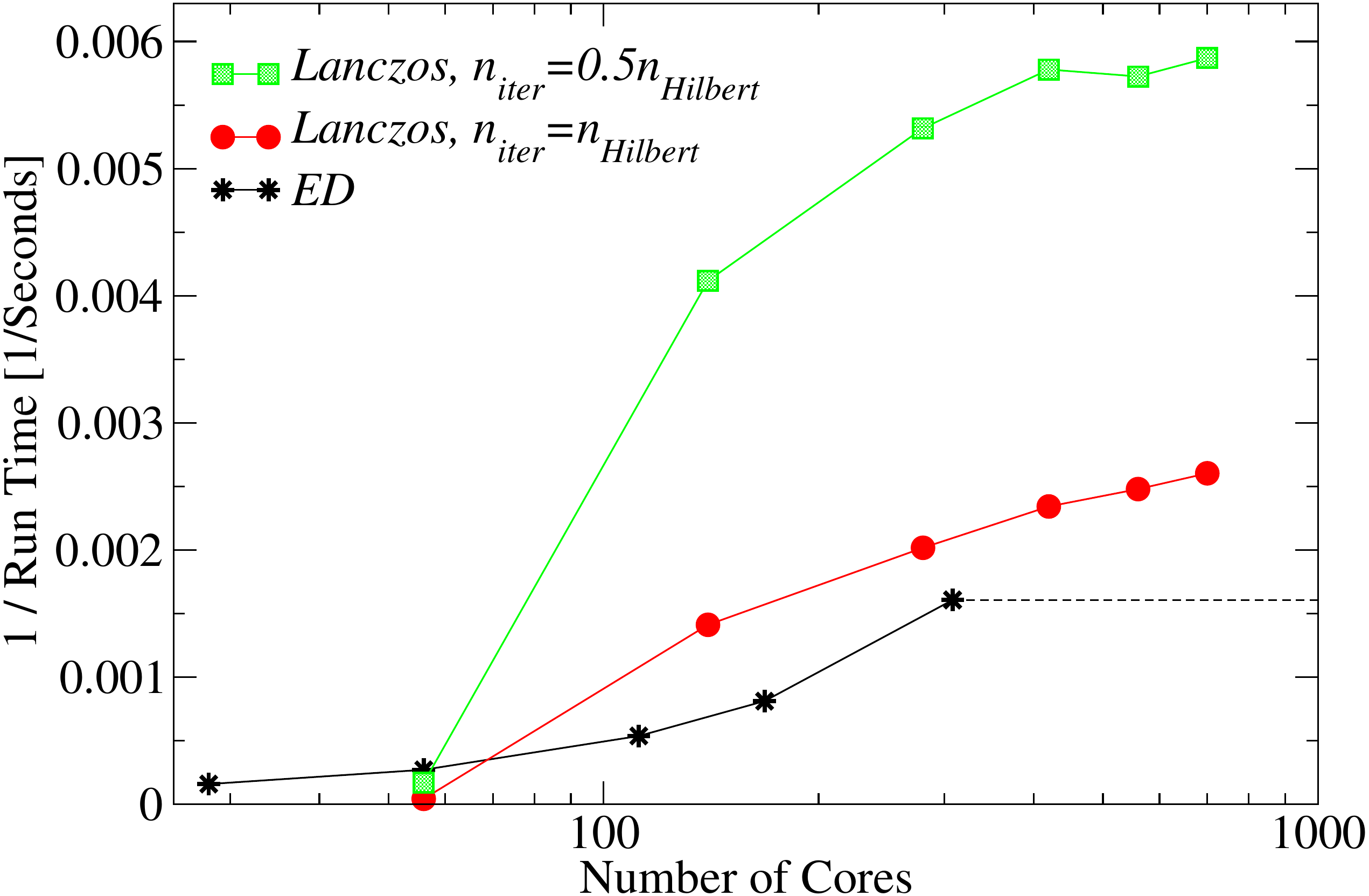}
\caption{Scaling of the Lanczos partial diagonalization algorithm with full re-orthogonalization for up to the $5$th order. 
For comparison, we also show the same scaling for ED for which we use 
the threaded version of Intel's MKL to diagonalize every cluster on a node with 28 cores.
The performance of ED improves rapidly by increasing the number of nodes until we have as many nodes as the 
number of clusters in the last order of the series. No speedup can be achieved beyond that. The Lanczos algorithm
with as $n_{iter}=n_{Hilbert}$ quickly surpasses ED in performance and continues to offer some speedup by increasing 
the number of cores. Choosing $n_{iter}=0.5n_{Hilbert}$ reduces the computational time by roughly a factor of three.
\label{fig:scaling}}
\end{figure}

We use message passing interface (MPI) parallelization for ED  
in our NLCE implementation. We assign a collection of clusters in a given order to a different 
node to be diagonalized. Every compute node in the high-performance computer cluster we have used for our calculations contains
128 gigabytes of random access memory (RAM) and 28 cores. Therefore, we speed up the diagonalization by using the 
threaded version of MKL and assigning all 28 cores of each node to diagonalize one cluster at a time with multiple
nodes simultaneously at work. In this picture, the run time is expected to be proportional to the number
of clusters assigned to each node for diagonalization, or inversely proportional to the number of nodes requested, in 
the limit of large number of clusters. In Fig.~\ref{fig:scaling}, we show the inverse run time (speedup) as a function of number of cores
($=28\times$ number of nodes) for ED in the fifth order. Since there are only 11 clusters in that order, the speed
generally increases with increasing the number of nodes, until we reach 11 nodes 
($308$ cores). Beyond that, there cannot be any improvement in the run time of ED.

We also use MPI to parallelize the inner loops of our Lanczos algorithm. Noting that the largest matrices 
we diagonalize in the fifth order have a linear size of 12870, and therefore, every core can store 
at least up to $n_{Hilbert}$ double-precision Lanczos vectors in the RAM, we distribute the computational tasks
for both the operation of the Hamiltonian on a Lanczos vector and the reorthogonalization step over all
cores available. For this reason, ideally the run time would be inversely proportional to the number of cores.
However, as shown in Fig.~\ref{fig:scaling}, we find that the overhead due to communication rapidly 
increases and the speedup eventually saturates. However, for $n_{iter}=n_{Hilert}$ our Lanczos-based method is already 
faster than our parallel ED scheme when using 5 nodes and remains larger than ED as the number 
of nodes increases. The reorthogonalization step scales like $n_{iter}^2\times n_{Hilbert}$ and is the most 
expensive part of our routine. Therefore, decreasing $n_{iter}$ can significantly speed up the 
calculations. We find that choosing $n_{iter}=0.5n_{Hilert}$, cuts the run time by a bout a factor of three
in the case of the fifth order as shown in Fig.~\ref{fig:scaling}.

In summary, we have shown that the Lanczos algorithm can be an efficient and flexible technique to partially 
diagonalize Hamiltonian matrices corresponding to the largest clusters in the NLCE, allowing one to access 
temperatures not previously accessible using Lapack full diagonalization routines. We benchmark our approach 
using the square expansion for the checkerboard lattice Heisenberg model with frustration and show that
by computing less than 50\% of the eigenvalues through a parallelized Lanczos algorithm in significantly 
less time than ED, we can reach temperatures relevant to the convergence regions of the energy and 
entropy of the model in the thermodynamic limit. 

More efficient implementations of the Lanczos algorithm, 
such as restart Lanczos or smart partial reorthogonalization, can be employed in the future to further 
reduce the computational cost. Our Lanczos approach can be generalized to produce eigenvectors in 
addition to eigenvalues, at an additional cost, for the calculation of other properties of interest. It can also be
employed for NLCE studies of other quantum lattice models in which the diagonalization steps sets the 
computational limit. Finally, highly-precise results from quantum Monte Carlo methods can also be used in 
the last order in a similar fashion to speedup the calculations, at least in cases where the calculations
are not hindered by the fermion sign problem.

\acknowledgements
We acknowledge Marcos Rigol for his early idea of employing the Lanczos algorithm for full diagonalization 
of clusters in the NLCE.
EK acknowledges support from the National Science Foundation (NSF) under Grant No. DMR-1609560. 
The computations were performed on the Spartan high-performance computing facility at San Jos\'{e}
State University supported by the NSF under Grant No.  OAC-1626645.

\end{document}